# Landau damping in hybrid plasmonics

*Alexander V. Uskov[1*], Jacob B. Khurgin,[2] Igor V. Smetanin,[1]*

*Igor E. Protsenko,[1] Nikolay V. Nikonorov[3]*

[1]P. N. Lebedev Physical Institute, Russian Academy of Sciences, Leninskiy Pr. 53, Moscow, 119333, Russia

[2]Department of ECE, John Hopkins University, Baltimore, Maryland 21218, United States

[3]ITMO University, Kronverskiy av. 49, St. Petersburg 197101, Russia

ABSTRACT. Landau Damping (LD) mechanism of the Localized Surface Plasmon (LSP) decay is studied for the hybrid nanoplasmonic (metal core/dielectric shell) structures. It is shown that LD in hybrid structures is strongly affected by permittivity and electron effective mass in the dielectric shell in accordance with previous observations by Kreibig, and the strength of LD can be enhanced by an order of magnitude for some combinations of permittivity and effective mass. The physical reason for this effect is identified as electron spillover into the dielectric where electric field is higher than in the metal and the presence of quasi-discrete energy levels in the dielectric. The theory indicates that the *transition absorption* at the interface metal-dielectric is a dominant contribution to LD in such hybrid structures. Thus, by judicious selection of dielectric material and its thickness one can engineer decay rates and hot carrier production for important applications, such as photodetection and photochemistry.

In the last few years [1], there has emerged a new research area, "hybrid plasmonics", whose subject is the structures comprising plasmonic metals, covered with thin layer of dielectric or semiconductor. These hybrid plasmonic structures are of great interest, and have many potential applications in photocatalysis, photovoltaics and photodetection [2-3]. All of these applications are based on photoexcitation of hot carriers in the metal with their subsequent injection into the dielectric layer and beyond. The excitation of hot carriers occurs via diverse mechanisms, such as phonon and defect –assisted absorption, absorption assisted by the electron-electron scattering, and, most important, the size-dependent absorption assisted by the metal/dielectric interface which on the quantum level can be construed as Landau damping (LD) [4-5]. In a broad sense of the word, LD refers to the optically-induced transition within conduction band in which the momentum conservation is assured by the presence of large spatial frequencies (wavevectors) in the spatial spectrum of the electro-magnetic mode. These high spatial frequencies are engendered by all kinds of non-uniformities and discontinuities of the electric field in the mode, and such discontinuities naturally occur at the interfaces. Thus, in Localized Surface Plasmon (LSP) mode, LD is associated with the surface of the metal nanoparticle and it has been also known as "Kreibig mechanism" (see [6] and references therein) given a simple phenomenological explanation of the process assisted by the electron collision with surface.

The LSP decay is known to manifest itself as broadening of dipole plasmonic resonance in metal nanoparticles, and since LD occurs at the surface, this broadening is size-dependent (1/R−law, "Kreibig broadening"), and is well-known for, at least, several decades [6]. Still, LD mechanism of broadening continues to attract much attention of both theoreticians and experimentalists [4-5, 7-17], since LD affects characteristics of nanophotonic devices [18], and measurement of Landau damping rates provides an effective tool in the physics of nanostructure surface [7, 16]. Most studies of LD have only considered simple structures – metal nanoparticles embedded into bulk dielectric or air while neglecting such intricacies as effective mass discontinuity.

In this Letter, we present a rigorous theoretical study of LD in hybrid plasmonic nanostructures. Using approach developed in [10], we demonstrate strong influence of the dielectric constant in the dielectric layer of hybrid structure on LD as had been previously found experimentally by Kreibig in [7]. This strong dependence of LD on ε is attributed, in particular, to the so-called *transition absorption* due to the discontinuity of the permittivity at the interface [19]. In addition, the discontinuity of electron effective mass at the metal-dielectric interface is also shown to affect LD.

The model that captures all the key features of hot carrier excitation in hybrid nanostructure is shown in Fig.1. We consider a metal nanoparticle with the dielectric constant $\varepsilon_m$, covered with a thin dielectric (semiconductor) of the thickness $d$ with the dielectric constant $\varepsilon_d$. The electron mass in metal $m_m \approx m_o$ where $m_o$ is the free electron mass, and in dielectric it is $m_d$. An electromagnetic wave of the frequency $\omega$ excites plasmonic oscillations in the nanoparticle. Because LD occurs at the metal surface the hot carriers excited via LD have the highest probability to leave the metal and perform useful function in detection, catalysis or other processes. The LD is described through an addition of a term $\Delta\gamma_{LD}$ to the damping rate $\gamma_c$ of electrons in bulk metal [6]: $\Delta\gamma_{LD} = A v_F / L_{nano}$, where $v_F$ is Fermi velocity in metal, $L_{nano}$ is the characteristic size of the nanoparticle. According to [10], the Kreibig coefficient $A$ is

$$A = A_{mat} A_{geom}, \tag{1}$$

where $A_{geom}$ depends on the nanoparticle size and shape, and on the spatial distribution of electric field in the nanoparticle (see also [10, 20]). For example, in the quasistatic approximation for dipole mode in spherical nanoparticle $A_{geom} = 1$ [20]. On the other hand, $A_{mat}$ depends on both frequency and material parameters, and in particular, on characteristics of the interface between metal nanoparticle and surroundings [10]:

$$A_{mat} = 0.5 \cdot \left(\omega^3 / \omega_p^2\right) \cdot \left(\hbar / \varepsilon_o v_F\right) \cdot K_R, \qquad (2)$$

where $\omega_p$ is the plasma frequency of the metal (we assume that $\varepsilon_m = 1 - \omega_p^2 / \left[\omega(\omega + i\gamma_c)\right]$), and $\varepsilon_o$ is the permittivity of vacuum. The coefficient $K_R$ determines the photon absorption rate $R$ per unit square of nanoparticle surface [1/(m$^2$·s)]: $R = K_R \cdot |F_m|^2$, where $F_m$ is the component of electric field in metal, *normal* to the interface of the nanoparticle. In [10] (see also [20]), the coefficient $A_{mat}$ had been calculated for the model with *infinite* potential barrier at the metal interface. Here, we calculate $A_{mat}$ for hybrid plasmonic nanostructures where the model with infinite barrier is not applicable.

In our calculations, the barrier height $W_d = 1$eV, and the work function $W_{vac}$ is approximated as $W_{vac} \to +\infty$ (the infinite barrier between dielectric and vacuum). Fermi energy in metal is taken as $\varepsilon_F = 5.5$eV. The normal components of electric field in metal $F_m$ and in dielectric $F_d$ satisfy the boundary condition $\varepsilon_m F_m = \varepsilon_d F_d$.

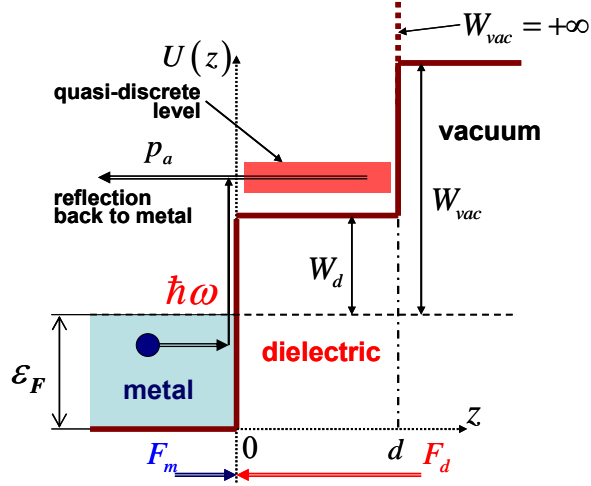

Fig. 1. Conduction band diagram of hybrid plasmonic structure − the electron potential $U(z)$ over the structure. Quasi-discrete level in dielectric layer is depicted with red bar. $F_m$ and $F_d$ are amplitudes of the optical field, normal to the interface, in metal and dielectric, correspondingly. Electron (blue circle) collides with the interface, acquires the photon energy $\hbar\omega$ from the field and reflected back to metal with the probability $p_a$. $\varepsilon_F$ is the Fermi energy in metal. $W_d$ and $W_{vac}$ are the barrier height and work functions correspondingly. $d$ is the thickness of dielectric layer. Dotted brown line illustrates the infinite barrier between dielectric and vacuum, used in calculations.

As a result of collisions with the interface, electrons in the metal can absorb a photon (LSP) $\hbar\omega$ and be reflected back to metal, as a" hot electron" [2-5]. The absorption rate coefficient $K_R$ in Eq.(2) is found by calculating quantum-mechanical probability $p_a$ for a *single* electron to absorb the energy $\hbar\omega$ during its collision with the interface, and then by summing over *all* electrons of metal colliding with the interface [10]. Using time-dependent perturbation theory in continuous spectrum [10, 21-24], we calculate the probability $p_a = p_a(\hbar\omega, E_{i,z}, E_{i,\parallel})$ for an electron in the metal with the initial energy $E_i = E_{i,z} + E_{i,\|}$, where $E_{i,z}$ ($E_{i,\|}$) is the kinetic energy of electron motion normal (parallel) to the interface following

$$p_a(\hbar\omega, E_{i,z}, E_{i,\|}) = (v_{f,z}/v_{i,z}) \cdot |C_-|^2 \qquad (3)$$

where $v_{i,z}$ ($v_{f,z}$) is the initial (final) velocity of electron in metal along axis $z$, and the amplitude $C_-$ is calculated as

$$C_- = \frac{ieF_m}{m_m \omega v_{f,z}} \cdot \left\{ \int_{-\infty}^{d} dz \frac{d\psi_i}{dz} \psi_{f,+} + \left( \frac{\varepsilon_m}{\varepsilon_d} \cdot \frac{m_m}{m_d} - 1 \right) \cdot \left[ \int_{0}^{d} dz \frac{d\psi_i}{dz} \psi_{f,+} + \frac{1}{2} \psi_i(0) \psi_{f,+}(0) \right] \right\} \qquad (4)$$

(see, for instance, Supporting Information SI-1). Correspondingly, the probability (3) can be written as

$$p_a = c_a(\hbar\omega, E_{i,z}, E_{i,\|}) \cdot |F_m|^2 \qquad (5)$$

where $c_a(\hbar\omega, E_{i,z}, E_{i,\|}) = (v_{f,z}/v_{i,z}) \cdot |C_-|^2 / |F_m|^2$. In Eq. (4), $\psi_i$ and $\psi_{f,+}$ are the electron wavefunctions in the initial state with the energies $E_{i,z}$ and $E_{i,\|}$, and in the final state with the energies $E_{f,z} = E_{i,z} + \hbar\omega$ and $E_{f,\|} = E_{i,\|} = E_\|$, respectively. Since $m_d \neq m_m$, the above wavefunctions must be found from 1D Schrödinger equation with the effective potential $U_{eff}(z) = U(z) + (1 - m_d/m_m) \cdot E_\|$ in dielectric ($0 < z < d$), depending on the energy $E_\|$ of electron motion parallel to the interface, and $U_{eff}(z) = U(z)$ in metal ($z < 0$). Thus, we have quasi-1D quantum-mechanical problem (see, more detail discussion in [24] and Supporting Information SI-1). Because in calculation, we approximate $W_{vac} \to +\infty$, the upper integration limit of $+\infty$ in integrals in Eq.(4) is replaced with $d$.

The first term in braces in Eq.(4) (referred as term I below) describes the absorption of photon (LSP) with energy $\hbar\omega$ due to collision of electron with the potential barrier (see Fig. 1). If $m_d = m_m$ and $U(z) \equiv 0$ (the barrier is absent), term I is equal to zero, and the second term (term II below), proportional to $[(\varepsilon_m/\varepsilon_d)-1]$ in this case, describes the pure *transition absorption* due to jump of the dielectric constant at the interface [19, 22]. If $U(z) \equiv 0$, $\varepsilon_d = \varepsilon_m$ and $m_d \neq m_m$, both terms in braces in Eq.(4) are nonzero, and they describe photon absorption due to jump of electron mass at the interface (and corresponding violation of translation symmetry in system) [22]. Of course, in real structures with nonzero barrier at the interface, $\varepsilon_d \neq \varepsilon_m$ and $m_d \neq m_m$, the complex amplitudes in Eq.(4) *interfere* with each other, and the picture, how the mechanisms of photon absorption are summed up, becomes nontrivial., Thus, this second term is responsible for the dependence of Landau damping on the dielectric constant of surroundings in plasmonic nanostructures – see below. Obviously, role of this term enhances with decreasing $\varepsilon_d$ and $m_d$ in Eq.(4).

Note that above approach to calculate the absorption probability $p_a$ has been used to evaluate the probability of electron *photoemission* in *surface* photoeffect [21-24].

Fig. 2 shows the spectrum of the probability coefficient $c_a(\hbar\omega, E_{i,z}, E_{i,\parallel})$ at given initial electron energies $E_{i,z}$=5eV, and $E_{i,\parallel}$=0. Black curve is for the infinite barrier, $W_d = \infty$, (as in [10, 20]), and it is considered as a reference. Solid curves take into account both two terms I and II in braces in Eq.(4); dotted (dashed) curve is calculated only with term I (term II). In red (blue) curves $m_d = 0.2m_o$ ($m_d = 0.5m_o$). Here and below we use the dielectric function of metal $\varepsilon_m$ with $\hbar\omega_p$=9eV, $\hbar\gamma_c$=0.07eV.

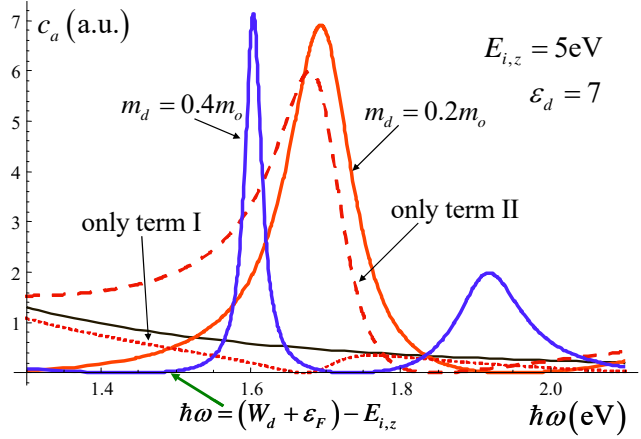

Fig. 2. Spectrum of the probability coefficient $c_a\left(\hbar\omega, E_{i,z}, E_{i,\|}\right)$ for $E_{i,z}$=5eV, $E_{i,\|}$=0. Black curve is for infinite barrier between metal and dielectric ($W_d = \infty$). In calculations, the thickness $d$ =3nm, $\varepsilon_d$ =7. Red curves: $m_d = 0.2 m_o$, blue curve: $m_d = 0.4 m_o$. Solid curves: both terms I and II in braces in Eq.(4) are taken into account; dotted (dashed) curves: only term I (II) works. Green arrow shows the photon energy $\hbar\omega = \left(W_d + \varepsilon_F\right) - E_{in,z}$ when electron in its final state reaches the bottom of the dielectric conduction band.

A few important observations can be made. First of all, one should note that the resonances in the curves are manifestation of the resonant quasi-levels emerging in conduction band of the dielectric shell. Increasing electron mass $m_d$, as well as increasing thickness $d$, result in the shift of quasi-levels closer to the bottom of conduction band, and accordingly, to red shifts of the resonances in the spectrum of $c_a$. Note, in blue curve in Fig. 2 resonances due to two quasi-levels are seen.

Furthermore, the resonances are all Fano-like as a consequence of "bound level in continuum" (see [25] and references therein), One can say that states of continuum in metal and the state in dielectric layer are coupled to each other *coherently,* since the states are described with *joint* wavefunction over whole structure in metal and dielectric. One can therefore talk also that "coherent electron transfer" occurs between metal and dielectric [26].

Finally, comparison of red curves shows that the contribution of the term II, which depends on the permittivities ratio $\varepsilon_m/\varepsilon_d$, dominates in braces in Eq.(4), leading to strong dependence of the LD on the dielectric constant $\varepsilon_d$ as demonstrated below.

To find $K_R$ in Eq.(2) for $A_{mat}$, one must sum over all electrons of metal, colliding with the interface of the metal [10]:

$$K_R = \frac{m_o}{2\pi^2\hbar^3} \int_0^\infty \int_0^\infty dE_{i,z} dE_{i,\|} \cdot c_a\left(\hbar\omega, E_{i,z}, E_{i,\|}\right) \cdot f_F\left(E_{i,z} + E_{i,\|}\right)\left[1 - f_F\left(E_{i,z} + E_{i,\|} + \hbar\omega\right)\right] \quad (6)$$

where $f_F(E_i)$ is the Fermi distribution of electrons in metal.

Fig. 3 presents the main result of this work. It shows the spectrum of the coefficient $A_{mat}$ for various effective masses $m_d$ and the dielectric constants $\varepsilon_d$ in dielectric layer. One can see that $A_{mat}(\hbar\omega)$ stays flat as long as $\hbar\omega < W_d$ but then it takes off rapidly when LSP energy $\hbar\omega$ exceeds barrier height $W_d$, and electron in the final state start spreading into the dielectric shell where the electric field is larger than in metal, $|F_d| > |F_m|$. This fact obviously facilitates LSP absorption at the interface, increasing $A_{mat}$. One can see that $A_{mat}$ increases with decreasing $m_d$ and $\varepsilon_d$ in accordance with Eq.(4). Comparison of dotted (term I) and dashed (term II) curves clearly demonstrates the dominant role of the contribution of transition absorption in $A_{mat}$.

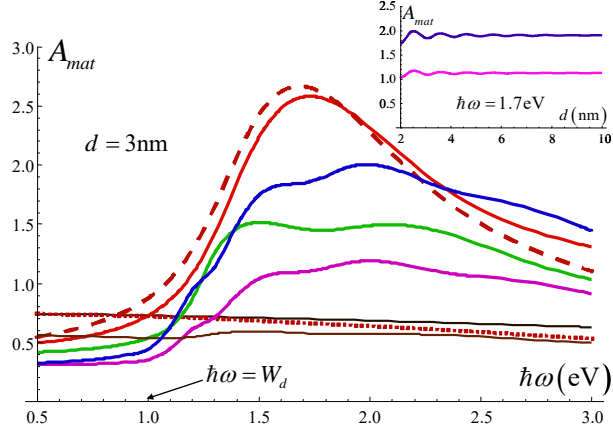

Fig. 3. Spectrum of the coefficient $A_{mat}$: $m_d = 0.1m_o$, $\varepsilon_d$=2.2 (red); $m_d = 0.5m_o$, $\varepsilon_d$=1.1 (blue); $m_d = 0.2m_o$, $\varepsilon_d$=2 (green); $m_d = 0.5m_o$, $\varepsilon_d$=1.5 (magenta); $m_d = 0.2m_o$, $\varepsilon_d$=7 (brown); black curve is for infinite barrier ($W = \infty$). Dotted (dashed) line takes into account only term I (II) in Eq.(4). In calculations, $d$=3nm. Arrow indicates the photon energy $\hbar\omega = W_d$. Inset: $A_{mat}$ as function of the thickness $d$ at given $\hbar\omega$=1.7eV, blue: $m_d = 0.5m_o$, $\varepsilon_d$=1.1; magenta: $m_d = 0.5m_o$, $\varepsilon_d$=1.5.

The inset in Fig. 3 illustrates the dependence of the coefficient $A_{mat}$ on the thickness $d$ of dielectric layer at given photon energy $\hbar\omega$=1.7eV for various $m_d$ and $\varepsilon_d$ in dielectric layer. The coefficient $A_{mat}$ exhibits small oscillations at low thicknesses, and becomes almost constant with increasing $d$. Such behavior of $A_{mat}$ with changing thickness $d$ concurs, with behavior of the plasmon damping in hybrid structures observed in [16].

Fig. 4 presents the coefficient $A_{mat}$ in the parameter space $(m_d, \varepsilon_d)$ at given $\hbar\omega$=1.7eV and $d$=3nm. Maximum of $A_{mat}$ in Fig.4 is 11.8 at $m_d$=0.1 and $\varepsilon_d$=1.

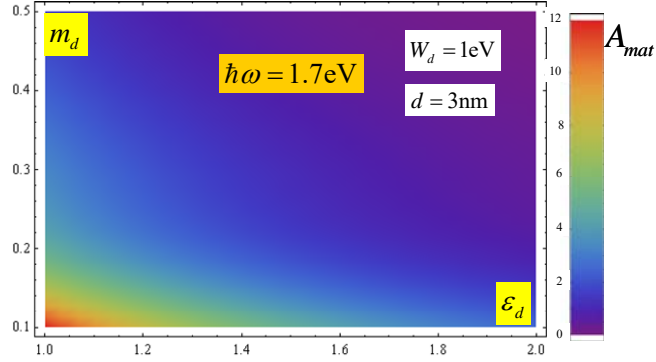

Fig. 4. Coefficient $A_{mat}$ as function of $\varepsilon_d$ and $m_d$. $\hbar\omega$=1.7eV, $d$=3nm.

In summary, the theory of LSP damping in hybrid plasmonic nanostructures developed here has led to the conclusion that LSP damping is strongly affected by both permittivity and effective mass of the dielectric layers. The main reasons for such dependence is electron penetration into the dielectric layer where electric field is stronger as well as the presence of quasi-bound resonant states in the dielectric. The latter phenomenon can be described as coherent transfer of carriers from metal into dielectric. The results obtained are in line with the earlier observations by Kreibig *et al* [7] as well as with more recent experiments [16].

Hot carriers excited via LD are very efficiently transferred out of metal. Otherwise, the hot carrier excited in the bulk tend to thermalize before reaching the interface. Therefore, the increase in LD means increase of the efficiency of such processes as photodetection and photocatalysis, and herein lies the main practical conclusion of this work.

Finally, it should also be noted that the dependence of Chemical Interface Damping (CID) [27-28] on the permittivity of material, surrounding nanoparticles, can also have the same origin as in hybrid plasmonics.

## ASSOCIATED CONTENT

**Supporting Information**

The derivation of formula for the amplitude $C_-$, Eq.(4).

## AUTHOR INFORMATION

**Corresponding Author**

**Alexander V. Uskov** – P. N. Lebedev Physical Institute, Russian Academy of Sciences, Leninskiy Pr. 53, Moscow, 119333, Russia; Email: uskovav@lebedev.ru

**Authors**

**Jacob B. Khurgin** – Department of Electrical and Computer Engineering, John Hopkins University, Baltimore, Maryland 21218, United States; Email: jakek@jhu.edu; https://orcid.org/0000-0003-0725-8736

**Igor V. Smetanin** – P. N. Lebedev Physical Institute, Russian Academy of Sciences, Leninskiy Pr. 53, Moscow, 119333, Russia; Email: smetaniniv@lebedev.ru

**Igor E. Protsenko** – P. N. Lebedev Physical Institute, Russian Academy of Sciences, Leninskiy Pr. 53, Moscow, 119333, Russia; Email: protsenk@gmail.com

**Nikolay V. Nikonorov** – ITMO University, Kronverskiy av. 49, St. Petersburg 197101, Russia; Email: nikonorov@oi.ifmo.ru

**Notes**

The authors declare no competing financial interest.

**Funding Sources**

Russian Science Foundation (20-19-00559).

ACKNOWLEDGMENT

A.U., I.S., I.P. and N.N. are thankful to the Russian Science Foundation (Grants 20-19-00559) for support.